\documentclass[twocolumn,epjc3]{svjour3}
\usepackage{graphics}
\usepackage{cuted}
\usepackage{flushend}

\usepackage{microtype}
\usepackage{bbm}
\usepackage{amsmath}
\usepackage{mathtools}
\usepackage{amssymb}
\usepackage{mathrsfs}       
\usepackage[pdftex]{graphicx}
\usepackage{pgfplots}
\pgfplotsset{width=10cm,compat=1.9}
\usepackage{xargs}       
\usepackage{wasysym}
\usepackage{engrec}
\usepackage{enumerate}
\usepackage{appendix}
\usepackage{empheq}
\usepackage{float}
\usepackage{stmaryrd}
\usepackage[parfill]{parskip}
\usepackage[pdftex,breaklinks]{hyperref}
\usepackage{titletoc}
\usepackage{makecell}
\usepackage{lscape}
\usepackage{algorithm}
\usepackage{algorithmicx}
\usepackage{tensor}
\usepackage{algpseudocode}
\usepackage{blkarray}
\usepackage{multirow}
\usepackage{bigstrut}
\usetikzlibrary{plotmarks}
\usepackage{braket}
\usepackage[export]{adjustbox}

\usepackage{simpler-wick}
\usepackage{bm}

\begin{document}
\title{\textit{Ab initio} description of monopole resonances in light- and medium-mass nuclei}
\subtitle{II. \textit{Ab initio} PGCM calculations in $^{46}$Ti, $^{28}$Si and $^{24}$Mg}
\author{A. Porro\thanksref{ad:tud,ad:emmi,ad:saclay} 
\and T. Duguet\thanksref{ad:saclay,ad:kul}
\and J.-P. Ebran\thanksref{ad:dam,ad:dam_u}
\and M. Frosini\thanksref{ad:cadarache}
\and R. Roth\thanksref{ad:tud,ad:darm2}
\and V. Som\`a\thanksref{ad:saclay}}

\institute{
\label{ad:tud}
Technische Universit\"at Darmstadt, Department of Physics, 64289 Darmstadt, Germany
\and
\label{ad:emmi}
ExtreMe Matter Institute EMMI, GSI Helmholtzzentrum f\"ur Schwerionenforschung GmbH, 64291 Darmstadt, Germany
\and
\label{ad:saclay}
IRFU, CEA, Universit\'e Paris-Saclay, 91191 Gif-sur-Yvette, France 
\and
\label{ad:kul}
KU Leuven, Department of Physics and Astronomy, Instituut voor Kern- en Stralingsfysica, 3001 Leuven, Belgium
\and
\label{ad:dam}
CEA, DAM, DIF, 91297 Arpajon, France
\and
\label{ad:dam_u}
Universit\'e Paris-Saclay, CEA, Laboratoire Mati\`ere en Conditions Extr\^emes, 91680 Bruy\`eres-le-Ch\^atel, France
\and
\label{ad:cadarache}
CEA, DES, IRESNE, DER, SPRC, 13108 Saint-Paul-l\`es-Durance, France
\and
%\label{ad:darm1}
%Institut f\"ur Kernphysik, Fachbereich Physik, Technische Universit\"at Darmstadt, Schlossgartenstr. 2, 64289 Darmstadt, Germany
%\and
\label{ad:darm2}
Helmholtz Forschungsakademie Hessen f\"ur FAIR, GSI Helmholtzzentrum, 64289 Darmstadt, Germany
}
\date{Received: \today{} / Revised version: date}

\maketitle
\begin{abstract}
Giant resonances (GRs) are a striking manifestation of collective motions in  atomic nuclei. The present paper is the second in a series of four dedicated to the use of the projected generator coordinate method (PGCM) for the \textit{ab initio} determination of the isoscalar giant monopole resonance (GMR) in closed- and open-shell mid-mass nuclei. 

While the first paper was dedicated to quantifying various uncertainty sources, the present paper focuses on the first applications to three doubly-open shell nuclei, namely $^{46}$Ti, $^{28}$Si and $^{24}$Mg. In particular, the goal is to investigate from an \textit{ab initio} standpoint (i) the coupling of the GMR with the giant quadrupole resonance (GQR) in intrinsically-deformed nuclei, (ii) the possible impact of shape coexistence and shape mixing on the GMR, (iii) the GMR based on shape isomers and (iv) the impact of anharmonic effects on the monopole response. The latter is studied by comparing PGCM results to those obtained via the quasi-particle random phase approximation (QRPA), the traditional many-body approach to giant resonances, performed in a consistent setting.

Eventually, PGCM results for sd-shell nuclei are in excellent agreement with experimental data, which is attributed to the capacity of the PGCM to capture the important fragmentation of the monopole response in light, intrinsically-deformed systems. Still, the comparison to data in $^{28}$Si and $^{24}$Mg illustrates the challenge (and the potential benefit) of extracting unambiguous experimental information. 

\end{abstract}

\section{Introduction}

The  present work is dedicated to the use of the projected generator coordinate method (PGCM) for the \textit{ab initio} determination of the isoscalar giant monopole resonance (GMR) in closed- and open-shell mid-mass nuclei. This represents the second paper of a series of four (Papers I-IV). The general motivations were detailed in the introduction to Paper I~\cite{Porro24a} and will not be repeated here. Specifically, Paper I was dedicated to quantifying several uncertainty sources in PGCM calculations of GRs. While the systematic and fully consistent evaluation of all uncertainty sources is a daunting task that can only be the result of a long-term effort, the main conclusion of Paper I was that \textit{ab initio} PGCM calculations of the GMR in mid-mass nuclei are pertinent and can be quantitative, at least when it comes to peaks carrying a large fraction of the strength.

In this context, Paper II focuses on the study of the GMR in three doubly-open shell nuclei, namely $^{46}$Ti, $^{28}$Si and $^{24}$Mg. The goal is to investigate from an \textit{ab initio} standpoint (i) the mechanism behind the coupling between the GMR and the giant quadrupole resonance (GQR) in intrinsically-deformed nuclei, (ii) the potential impact of shape coexistence and shape mixing effects on the GMR, (iii) the monopole response on top of shape isomers and (iv) the impact of anharmonic effects on the monopole response in light nuclei. Results in $^{28}$Si and $^{24}$Mg are confronted with the outcome of recent experimental measurements.

Paper II is organised as follows. After recalling the computational set up in Sec.~\ref{sec:numerical}, Sec.~\ref{sec:O16_resphys} provides a reference calculation of the monopole response of $^{16}$O, which acts as an archetypal spherical, doubly-closed-shell system. The physics of the GMR-GQR coupling in PGCM calculations is discussed at length in Sec.~\ref{sec:Ti46_resphys} for $^{46}$Ti. Next, the potential impact of shape-coexistence and shape-mixing effects on the GMR in $^{28}$Si are addressed in Sec.~\ref{sec:Si28_resphys}, whereas the highly-fragmented monopole response in $^{24}$Mg is investigated in Sec.~\ref{sec:Mg24_resphys}. Eventually, a comparison between PGCM and QRPA calculations performed in a consistent setting is provided in Sec.~\ref{sec:QFAM_res} to evaluate the degree of anharmonicity at play. While conclusions are provided in Sec.~\ref{sec:concl}, an Appendix details the schematic model used to quantify the impact of anharmonic effects on the GMR.

\section{Numerical aspects}
\label{sec:numerical}

The PGCM formalism and the setting of the numerical applications were laid down in details in Paper I. Only essential features about the latter are repeated here.

All calculations presented in Paper II use a one-body spherical harmonic oscillator basis characterised by the optimal frequency $\hbar\omega=12$~MeV. All states up to $e_{\!_{\;\text{max}}}\equiv\text{max}(2n+l)=10$ are included, with $n$ the principal quantum number and $l$ the orbital angular momentum. The representation of three-body operators is further restricted, due to computational limitations, by only employing three-body states  up to $e_{\!_{\;\text{3max}}}=14$.

\begin{figure}[b]
    \centering
    \includegraphics[width=\columnwidth]{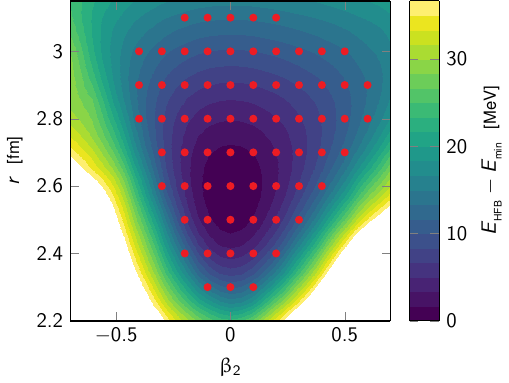}
    \caption{Two-dimensional HFB total energy surface in $^{16}$O relative to its minimum. The red dots correspond to the set of constrained HFB states included in the PGCM ansatz.}
    \label{fig:O16_PES}
\end{figure}

A chiral effective field theory ($\chi$EFT) Hamiltonian built at next-to-next-to-next-to-leading-order (N$^3$LO)~\cite{Hu20} is employed and contains consistent two- (2N) and three-nucleon (3N) interactions. The Hamiltonian is further evolved via similarity renormalisation group (SRG) transformation~\cite{Bogner:2009bt} to the low-momentum scale $\lambda=1.88$~fm$^{-1}$ (i.e. flow parameter $\alpha$=0.08~fm$^4$). Three-body forces are approximated via the rank-reduction method developed in Ref.~\cite{Frosini21a}.

Two-dimensional (2D) PGCM calculations mix constrained HFB states with axial symmetry using the root-mean-square $r$ and the axial mass quadrupole deformation parameter $\beta_2$ as generator coordinates. This means that PGCM calculations are presently restricted to accessing the $K=0$ component of the quadrupole strength function. The QRPA is performed at the HFB minimum employing the quasi-particle finite amplitude method (QFAM)~\cite{Beaujeault23a}.

\section{$^{16}$O: an archetypal spherical system}
\label{sec:O16_resphys}

\begin{table}[b]
    \centering
    \begin{tabular}{lcr}
    	\hline
        $E_{\;\text{HFB}}$ [MeV] & $r$ [fm] & $\beta_2$ \\
        \hline
        -79.55 & 2.580 & 0.00 \\ 
        \hline
    \end{tabular}
    \caption{Total energy, rms radius and quadrupole deformation $\beta_2$ of the HFB minimum in $^{16}$O.}
    \label{tab:O16_prop}
\end{table}
\begin{figure}[b]
    \centering
    \includegraphics[width=\columnwidth]{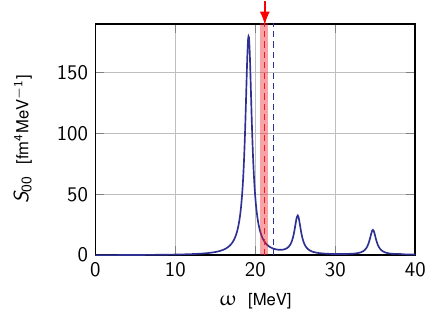}
    \caption{Monopole response in $^{16}$O obtained from the PGCM based on the mixing of  HFB states displayed in Fig.~\ref{fig:O16_PES}. The dashed blue and red lines represent the PGCM and experimental~\cite{Lui01a} centroids, respectively. The shaded area indicates the experimental uncertainty.}
    \label{fig:O16_spectra}
\end{figure}

In order to provide a baseline for the subsequent study, the monopole response in the intrinsically spherical, doubly closed-shell $^{16}$O nucleus is addressed first.

The 2D HFB total energy surface (TES) displayed in Fig.~\ref{fig:O16_PES} demonstrates that $^{16}$O  displays a well-defined spherical minimum at the mean-field level and a rather stiff topology with respect to both $r$ and $\beta_2$. The main characteristics of the HFB minimum are provided in Tab.~\ref{tab:O16_prop}. Based on the TES, a set of HFB states (red points) is selected for the subsequent PGCM calculation.

The corresponding monopole response shown in Fig.~\ref{fig:O16_spectra} displays a single GMR peak at $19.2$~MeV, followed by smaller peaks at about $25$ and $35$~MeV. The corresponding centroid at 21.13 MeV (dashed blue line) compares well with the experimental value at 22.26 MeV~\cite{Lui01a} (dashed red line).

\begin{figure}
    \centering
    \includegraphics[width=\columnwidth]{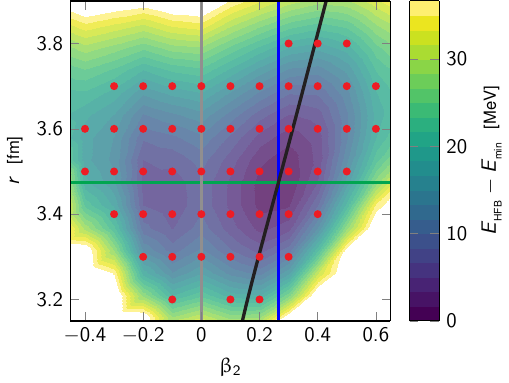}
    \caption{Two-dimensional HFB total energy surfance in $^{46}$Ti relative to its minimum. The red dots correspond to the constrained HFB states mixed in the PGCM ansatz. Coloured lines represents the different one-dimensional cuts used in the PGCM analysis, see text for details. The same colour code is used in Fig.~\ref{fig:spectra_Ti46_comp}.}
    \label{fig:PES_Ti46}
\end{figure}

\section{Deformation effects in $^{46}$Ti}
\label{sec:Ti46_resphys}

The 2D HFB TES in $^{46}$Ti is displayed in Fig.~\ref{fig:PES_Ti46}. The minimum, whose main characteristics are given in Tab.~\ref{tab:Ti46_prop}, is prolate  with the rather large deformation $\beta_2=0.27$. This nucleus is also predicted to display an oblate shape isomer located about $6.1$~MeV above the prolate absolute minimum with $\beta_2=-0.15$ and a similar radius. Based on the TES, a set of HFB states (red points) is selected for the subsequent 2D PGCM calculation.

\begin{figure}
    \centering
    \includegraphics[width=\columnwidth]{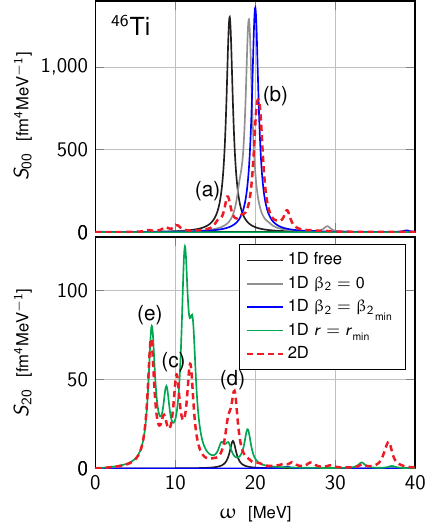}
    \caption{Monopole (top) and ($K=0$) quadrupole (bottom) responses in $^{46}$Ti for the different sets of PGCM calculations defined through Fig.~\ref{fig:PES_Ti46}. Labels (a) - (d) identify different peaks of the spectra discussed in the text.}
    \label{fig:spectra_Ti46_comp}
\end{figure}

The corresponding monopole and ($K=0$ component of the) quadrupole responses are shown in Fig.~\ref{fig:spectra_Ti46_comp}. Compared to $^{16}$O where the GMR is located into a single narrow peak, the monopole strength function is fragmented into three peaks: a small peak at $16.5$~MeV (peak (a)), a large peak at $20.4$~MeV  (peak (b)) followed by another small peak at about $24$~MeV. The $K=0$ component of the quadrupole strength function behaves differently with a first dominant low-energy peak at $7.0$~MeV (peak (e)) accompanied by three peaks at $8.6$, $10.2$ (peak (c)) and $11.9$~MeV. Next comes a distinct peak at $17.1$~MeV (peak (d)). 

\begin{figure*}
    \centering
    \includegraphics[width=\textwidth]{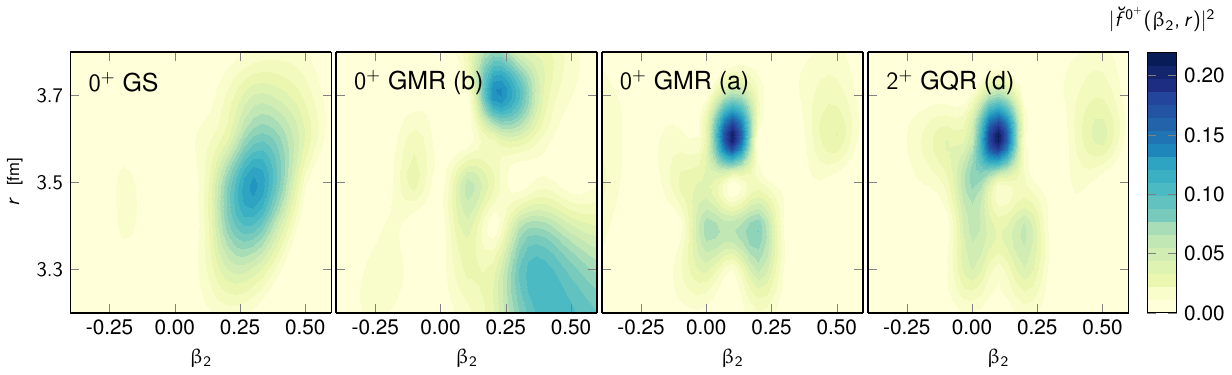}
    \caption{Intrinsic PGCM collective wave-functions in the $(r,\beta_2)$ plane of several states of interest (labels refer to Fig.~\ref{fig:spectra_Ti46_comp}) in $^{46}$Ti: (left) $0^+$ ground state, (second left) main $0^+$ component (a) of the GMR, (second right) lower-energy $0^+$ component (b) of the GMR, (right) $2^+$ GQR (d).}
    \label{fig:Ti46_WF_PGCM}
\end{figure*}

\begin{figure}
    \centering
    \includegraphics[width=\columnwidth]{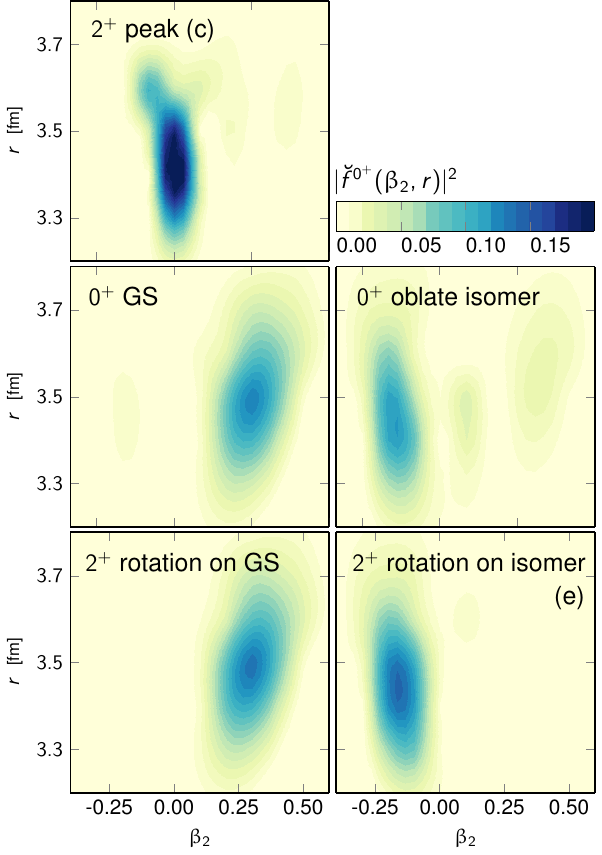}
    \caption{Intrinsic PGCM collective wave-functions in the $(r,\beta_2)$ plane of several states of interest (labels refer to Fig.~\ref{fig:spectra_Ti46_comp}) in $^{46}$Ti: (top left) $2^+$ state (c), (middle left) $0^+$ ground state, (bottom left) $2^+$ state belonging to the ground-state rotational band, (middle right) $0^+$ oblate-shape isomer, (bottom right) $2^+$ state (e) belonging to the oblate-shape isomer rotational band.}
    \label{fig:Ti46_WF_PGCM_iso}
\end{figure}

The objective is now to interpret several key patterns of the 2D PGCM strength functions just identified. To do so, two tools are employed. First, four specific one-dimensional (1D) reduced PGCM calculations are performed in connection with the 4 coloured lines appearing in Fig.~\ref{fig:PES_Ti46}
\begin{enumerate}
\item Fixed deformation
\begin{itemize}
\item Gray line: constrained $r$ at fixed $\beta_2=0$
\item Blue line: constrained $r$ at fixed $\beta_2=(\beta_2)_{\text{min}}$
\end{itemize}
\item Fixed radius
\begin{itemize}
\item Green line: constrained $\beta_2$ at fixed $r=(r)_{\text{min}}$
\end{itemize}
\item Adaptive deformation
\begin{itemize}
\item Black line: constrained $r$ with free $\beta_2=\beta_2(r)$
\end{itemize}
\end{enumerate}
The corresponding strength functions are also displayed in Fig.~\ref{fig:spectra_Ti46_comp}. Second, the intrinsic collective wave-functions from the 2D PGCM calculation corresponding to the labelled peaks in Fig.~\ref{fig:spectra_Ti46_comp} are shown in Figs~\ref{fig:Ti46_WF_PGCM} and~\ref{fig:Ti46_WF_PGCM_iso} to better interpret the nature of the states at play.

\begin{table}
    \centering
    \begin{tabular}{llcr}
    	\hline
        ~ & $E_{\;\text{HFB}}$ [MeV] & $r$ [fm] & $\beta_2$ \\
        \hline
        $^{46}$Ti & -237.30 & 3.474 & 0.27 \\ 
        $^{46}$Ti$_{\;\text{iso}}$ & -231.85 & 3.440 & -0.15 \\
        \hline
    \end{tabular}
    \caption{Total energy, rms radius and quadrupole deformation $\beta_2$ of the HFB minimum and oblate isomer in $^{46}$Ti.}
    \label{tab:Ti46_prop}
\end{table}

\subsection{Pure monopole vibrations (breathing mode)}

The main peak (b) in the monopole strength function of $^{46}$Ti is a reminiscence of the GMR in spherical nuclei where the intrinsic deformation plays no role, e.g. the single peak seen in $^{16}$O. This is confirmed by comparing the associated intrinsic PGCM collective wave-function in Fig.~\ref{fig:Ti46_WF_PGCM} (second from the left) to the one of the $0^+$ ground state (first from the left). The wave function of peak (b) displays a node in the radial direction at the ground-state deformation. Thus, this state exhibits the textbook picture of a pure breathing mode built as a radial vibration on top of the (intrinsically-deformed) ground-state. 

The above analysis is validated by focusing on the blue curve in the upper panel of Fig.~\ref{fig:spectra_Ti46_comp} associated with the 1D PGCM calculation realised by forcing the deformation $\beta_2=(\beta_2)_{\text{min}}$ to remain equal to the ground-state one throughout the radial vibration. The corresponding monopole strength function associated with pure radial oscillations at fixed deformation contains a single peak that is almost superposed with peak (b) from the 2D PGCM calculation, i.e. it is located only $400$~keV below it\footnote{Notice that in intrinsically-spherical nuclei such as  $^{16}$O, the mixing of deformed HFB states into the PGCM ansatz does not impact the monopole response, i.e. results from the 2D calculation are strictly equal to those from the 1D calculation obtained by fixing the deformation to $\beta_2=(\beta_2)_{\text{min}}=0$.}. Peak (a) is simply absent from such a 1D PGCM calculation. Repeating the same 1D PGCM calculation at zero fixed deformation (gray line) further shifts the peak down by $800$~keV, i.e. at $19.2$~MeV\footnote{Interestingly, it is exactly at the same energy as the GMR in $^{16}$O. Empirically, the GMR is known to scale as $82\cdot A^{-1/3}$~\cite{Ring80a}, i.e. which gives $32.5$~MeV in $^{16}$O and $22.9$~MeV in a spherical nucleus of mass $A=46$. Obviously, this empirical law does not work well in light nuclei. Furthermore, while this empirical law basically indicates how the radial curvature evolves with $A$ in spherical nuclei, the radial curvature at zero deformation in deformed nuclei has no reason to evolve similarly.}. The shift of $+800$~keV isolates the impact of the {\it static} intrinsic ground-state deformation on the spherical-like GMR peak. It can be understood by comparing the curvatures along the radial coordinate of the two 1D cuts at $\beta_2=0$ and $\beta_2=(\beta_2)_{\text{min}}$. The more pronounced curvature of the latter, especially against compression, indeed pushes the resonance up. This behavior is in qualitative agreement with the fluid-dynamical model of Ref.~\cite{Nishizaki85a}, where an increase of the deformation parameter induces a shift up in the energy of the resonance. The further increase by $+400$~keV to reach peak (b) in the 2D calculation corresponds to the additional impact of {\it dynamical} deformation, i.e. quadrupole fluctuations, on the spherical-like GMR peak.  

\subsection{Pure quadrupole vibrations}
\label{sec:Ti46_res_quad}

Pure quadrupole vibrations are obtained via the 1D PGCM calculation realised by forcing the radius $r=(r)_{\text{min}}$ to remain equal to the ground-state one. The associated results correspond to the green line in the lower panel of Fig.~\ref{fig:PES_Ti46}. While peak (d) at $17.1$~MeV is not accounted for, the low-lying states below $15$~MeV are qualitatively (and in some cases quantitatively) reproduced. Starting from this ascertainment, the nature of states (c) and (e) emerging from this 1D calculation is now briefly elaborated on in more details. 

The intrinsic PGCM wave-function of the $2^+$ state (c) is depicted in Fig.~\ref{fig:Ti46_WF_PGCM} (upper left) and is characterised by a sharp displacement along $\beta_2$ compared to the ground state without any structure along the radial direction. The wave function is sharply located around $\beta_2=0$ and probably corresponds to a state that could be described as a low-rank two quasi-particle (one-particle/one-hole) excitation on top of the spherical point. 

The nature of the $2^+$ state corresponding to the lowest-lying transition\footnote{The transition to the $2^+$ state at $510$~keV belonging to the ground-state rotational band,  whose collective PGCM intrinsic wave-function is shown in Fig.~\ref{fig:Ti46_WF_PGCM} (bottom left), has been removed from the quadrupole strength function in Fig.~\ref{fig:PES_Ti46}. Indeed, displaying the transition  $|\braket{2^+_{\!_\text{\;GSrot}}|Q_{20}|0^+_{\!_\text{\;GS}}}|^2=698.95$ fm$^4$MeV$^{-1}$ would completely squeeze down the rest of the strength function.} at $7.0$~MeV is different. In the monopole channel, there exists a $0^+$ state at $6.6$~MeV whose PGCM intrinsic wave-function displayed in Fig.~\ref{fig:Ti46_WF_PGCM_iso} (top left) is localised within the well of the oblate minimum of the 2D HFB TES (Fig.~\ref{fig:PES_Ti46}). This state is the oblate-shape isomer whose excitation energy is indeed consistent with the energy difference ($6.45$ MeV; see Tab.~\ref{tab:Ti46_prop}) between the oblate and prolate minima in the HFB TES. As seen in the upper panel of Fig.~\ref{fig:PES_Ti46}, the monopole strength (and thus the $E0$ transition strength to the ground state) is very small because of the absence of mixing between the two wells as testified by the ground-state intrinsic PGCM wave function in Fig.~\ref{fig:Ti46_WF_PGCM}.  Given this $0^+$ shape isomer, the $2^+$ state (d) is nothing but the rotational excitation built on top of it and located $380$~keV above the band-head. Its intrinsic PGCM wave-function displayed in Fig.~\ref{fig:Ti46_WF_PGCM_iso} (bottom right) is essentially identical to the $0^+$ shape isomer. Because the ground-state and shape isomer radii are very close (see Tab.~\ref{tab:Ti46_prop} for the HFB values), the rotational state built on the latter can be obtained via a pure (projected) quadrupole excitation of the former at fixed radius as shown by the 1D calculation in the bottom of Fig.~\ref{fig:PES_Ti46}. The fact that the $2^+$ state (e) is a rotational excitation of the shape isomer is confirmed by the strong $B(E2)[0^+_{\text{iso}}\rightarrow 2^{+}_{\text{(e)}}]=186.37$~fm$^4$MeV$^{-1}$ linking the two states.

\subsection{Coupling to the GQR}
\label{sec:coupling_Ti46}

The low-energy component (a) at $16.5$~MeV of the GMR is absent from pure radial 1D PGCM calculations at fixed deformation. At the same time, the GQR peak (d) at $17.1$~MeV does not emerge from pure quadrupole 1D PGCM calculations at fixed radius. Hence, these two states seem to rely on a collective motion involving both $r$ and $\beta_2$. 

This intuition is validated in Fig.~\ref{fig:PES_Ti46} via the results (black lines) from the 1D PGCM calculation constraining $r$ while leaving $\beta_2$ free to adjust. In paper I, $\beta_2$ was shown to be linearly correlated with $r$ in such a case. Eventually, the monopole (quadrupole) response from the 1D calculation contains a single peak at $16.8$~MeV ($17.2$~MeV) clearly capturing the low-energy component (a) of the GMR (GQR peak (d)). Going beyond the strict linear correlation beyond $r$ and $\beta_2$, the more elaborate 2D PGCM calculation slightly shifts the $0^+$ state down by $300$~keV while decreasing the associated monopole transition strength. At the same time, the $2^+$ state gets fragmented and the associated quadrupole transition strength is increased.

\begin{figure}
    \includegraphics[scale=0.97,right]{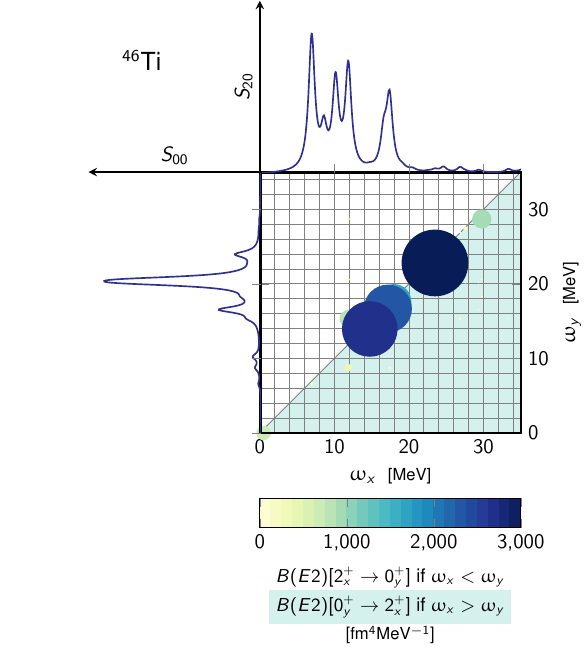}
    \caption{Upward $B(E2)$ transitions between all $0^+$ and $2^+$ states issued from the two-dimensional PGCM calculation of $^{46}$Ti. The dots are located at the intersection of the energies between $0^+$ (vertical axis) and $2^+$ (horizontal axis) states. The monopole (quadrupole) response of the ground state is also shown for comparison on the vertical (horizontal) axis. The size of the dots and the colour coding both reflect the magnitude of the transition.}
    \label{fig:Ti46_PP_coupling}
\end{figure}

The collective intrinsic wave-function of the $0^+$ state (a) in Fig.~\ref{fig:Ti46_WF_PGCM} (second from the right) displays a characteristic radial oscillation, like state (b) analyzed earlier on. 
The one of state (a) is however located at smaller deformation ($\beta_2\sim0.1$) than the ground state, thus implying an additional fluctuation in the $\beta_2$ direction. The presence of peak (a) in the monopole response is thus said to be due to the coupling of the GMR with the $K=0$ component of the GQR in intrinsically-deformed systems~\cite{Peru08a,Gambacurta20a}. In addition to appearing at nearly the same energy as the GQR peak (d), the intrinsic collective wave function of state (a) is strikingly similar to the intrinsic collective wave function of the $2^+$ state making up that GQR (first from the right). The present analysis nicely demonstrates that the PGCM generates these two states by projecting on $J^\pi =0^+$ and $2^+$ the same intrinsic collective state containing fluctuations in both the $r$ and $\beta_2$ directions. The capacity of the PGCM to do so and account {\it at mean-field like cost} for the physics at play in the GMR-GQR coupling {\it within a symmetry-conserving framework} strongly plays in favor of the method.

\subsection{$B(E2)$ transition probabilities}
\label{sec:E2}

Upward $B(E2)$ transition probabilities between $0^+$ and $2^+$ states predicted by the PGCM in $^{46}$Ti are now analysed. In Fig.~\ref{fig:Ti46_PP_coupling} \textit{all} such transitions are displayed, the points' size being proportional to their magnitude. Only few dominant transitions concentrated in the GQR region dominate and are eventually visible on the graph. Being located close to the diagonal, these dominant transitions connect $0^+$ and $2^+$ states at nearly the same energies and are much larger than the transitions in the ground-state quadrupole strength function. 

A large transition is seen to connect the low component of the GMR peak at 16.5 MeV and the GQR at 17.1 MeV,  $|\braket{2^+_{\!_\text{(d)}}|Q_{20}|0^+_{\!_\text{(a)}}}|^2=294$ fm$^4$MeV$^{-1}$. This transition is of the same order as the expectation value of the quadrupole operator in the GQR state, $|\braket{2^+_{\!_\text{(d)}}|Q_{20}|2^+_{\!_\text{(d)}}}|^2=464$ fm$^4$MeV$^{-1}$. This is the fingerprint that, as analysed above, the two states share the same intrinsic collective wave-function as a result of the GMR-GQR coupling mechanism. 

One observes two other extremely large $B(E2)$ transition probabilities connecting  $0^+$ and $2^+$ states sharing the same collective intrinsic wave function. Interestingly, the lowest one in energy connects a $0^+$ state and a $2^+$ state that are barely connected to the ground state. The second one involves the third peak of the GMR at about $24$~MeV.

\begin{figure}
    \centering
    \includegraphics[width=\columnwidth]{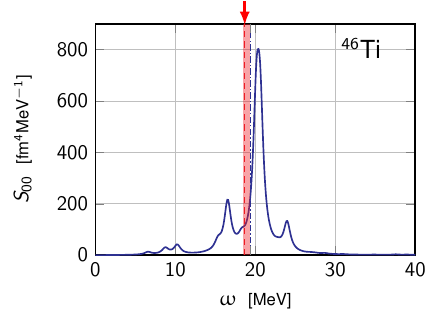}
    \caption{Monopole response in $^{46}$Ti for PGCM. The PGCM ansatz results from the combination of the points displayed in Fig.~\ref{fig:PES_Ti46}. The dashed blue and red lines represent, respectively, the PGCM and experimental~\cite{Tokimoto06a} centroid. Shaded areas indicate experimental uncertainties.}
    \label{fig:spectra_Ti46_exp}
\end{figure}

\subsection{Comparison to experiment}

The PGCM monopole response is eventually compared in Fig.~\ref{fig:spectra_Ti46_exp} to available experimental data, which are limited to the centroid value~\cite{Tokimoto06a} in the present case. The PGCM centroid (19.40~MeV) is shown to agree with the experimental (18.66~MeV) value (red dashed line) within uncertainties\footnote{As all theoretical predictions contained in the present study, this theoretical centroid must of course be accompanied by the associated uncertainties (partly) evaluated in Paper I.}.

\section{Shape coexistence in $^{28}$Si}
\label{sec:Si28_resphys}

\begin{figure}
    \centering
    \includegraphics[width=\columnwidth]{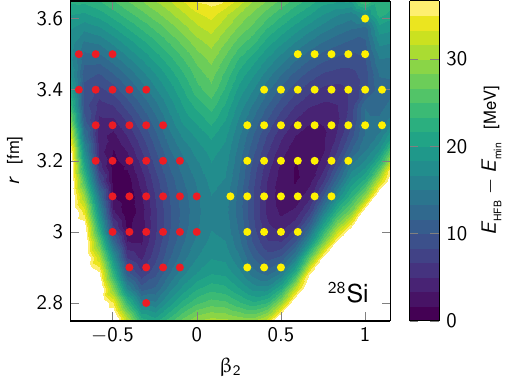}
    \caption{HFB total energy surface for $^{28}$Si in the $(\beta_2,r)$ plane. The red (yellow) dots correspond to the oblate (prolate) configurations included in the PGCM ansatz.}
    \label{fig:PES_Si28}
\end{figure}

Results for $^{28}$Si are now discussed. While the ground-state of this nucleus is known to be oblate, a prolate-shape isomer has been both predicted theoretically \cite{Darai12a} and observed experimentally \cite{Kelly98a,Jenkins12a}. This is presently confirmed by the 2D HFB TES displayed in Fig.~\ref{fig:PES_Si28}. Two distinct minima are observed, the lowest being associated to the oblate ground state ($\beta_2\approx-0.45$) and the second producing a prolate-shape isomer ($\beta_2\approx0.55$). The two HFB states have comparable radii as can be seen from Tab.~\ref{tab:Si28_prop}. While the energy of the two HFB minima differ by less than 1~MeV, they are however separated by a large barrier culminating at $16.5$~MeV above the oblate minimum. 

\begin{figure}
	\centering
	\includegraphics[width=\columnwidth]{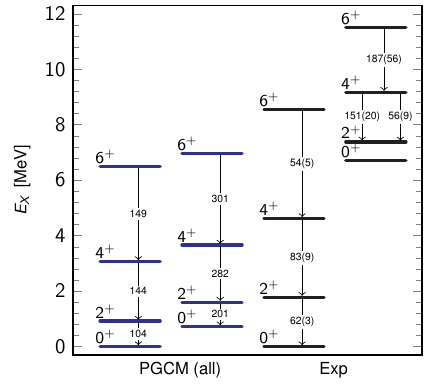}
	\caption{Rotational bands in $^{28}$Si relative to the ground state and the prolate-shape isomer. Associated $B(E2)$ values are expressed in  fm$^4$MeV$^{-1}$. Experimental data from \cite{Jenkins12a}.}
	\label{fig:LL_Si28_both}
\end{figure}

\begin{table}
    \centering
    \begin{tabular}{llcr}
    	\hline
        ~ & $E_{\;\text{HFB}}$ [MeV] & $r$ [fm] & $\beta_2$ \\
        \hline
        $^{28}$Si & -135.89 & 3.107 & -0.45 \\ 
        $^{28}$Si$_{\;\text{iso}}$ & -134.41 & 3.168 & 0.60 \\
        \hline
    \end{tabular}
    \caption{Total energy, rms radius and quadrupole deformation $\beta_2$ of the oblate and prolate (isomer) HFB minima of the $^{28}$Si TES from Fig.~\ref{fig:PES_Si28}.}
    \label{tab:Si28_prop}
\end{table}

The objective is to study the influence of shape coexistence and the possible shape mixing on the monopole response, knowing that the PGCM is well positioned to do so. To quantitatively address this question, three distinct 2D PGCM calculations are performed: one only including oblate HFB states (red points), one only including prolate HFB states (yellow points) and a third one including both oblate and prolate HFB configurations. A colour code is presently adopted to display the strength functions: blue is associated to the results of the full 2D PGCM calculation, whereas red (yellow) refers to the PGCM calculation restricted to oblate (prolate) configurations.

\subsection{Rotational bands}

Before coming to the responses, the rotational bands built on the ground state and on the prolate isomer are briefly studied. They are both displayed in Fig.~\ref{fig:LL_Si28_both} and compared to available experimental data. Data relative to the direct decay from the $2^+$ to the supposed $0^+$ isomer are absent. In fact, two very close $2^+$ states were observed, which explains the presence of two decays from the $4^+$ state. 

The computed ground-state rotational band is reasonable but is more compressed than the experimental one. The opposite is true for the prolate-shape isomer even though the discrepancy is less pronounced. The predicted intra-band $B(E2)$ are systematically too large by a factor of two. 

The main difference to the data, however, relates to the excitation energy of the isomeric band-head that is predicted to be $\sim6$~MeV lower than in the experiment. This feature can only artificially increase the effect of the shape mixing, if any, in the theoretical calculation.

\begin{figure}[t]
    \centering
    \includegraphics[width=\columnwidth]{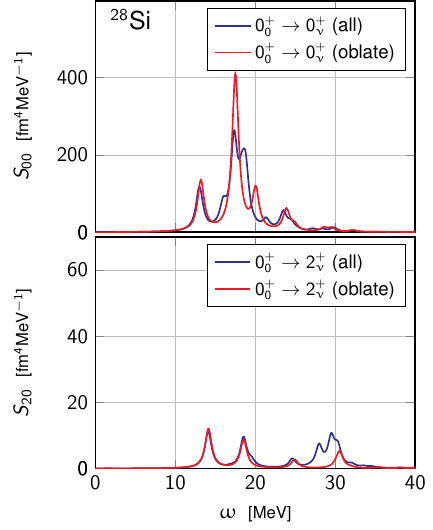}
    \caption{Monopole (top panel) and quadrupole (bottom panel) response of the oblate ground state of $^{28}$Si. The red (blue) curve refers to the PGCM calculation limited to oblate (including both oblate and prolate) configurations displayed  in Fig.~\ref{fig:PES_Si28}.}
    \label{fig:spectra_Si28_obl}
\end{figure}

\subsection{Ground-state responses}

The monopole and quadrupole responses relative to the ground state are displayed in Fig.~\ref{fig:spectra_Si28_obl}. A giant monopole resonance is identified at $\sim18$ MeV, accompanied by minor peaks both at lower and higher energies. The main GMR peak is slightly more fragmented in the full calculation due to the coupling to prolate configurations. Still, the results of the two calculations are very similar, showing that the impact of the shape mixing on the GMR is rather small. This feature, despite the (dubious) proximity in energy of both minima, is probably due to the large barrier separating both wells.

The quadrupole response displays two well-separated peaks that can be associated with the first two peaks in the monopole response. Some additional strength appears in the $[25,35]$~MeV interval for both calculations, with the coupling between the two wells isolating it in two separate peaks.

\subsection{Shape-isomer responses}

\begin{figure}[t]
    \centering
    \includegraphics[width=\columnwidth]{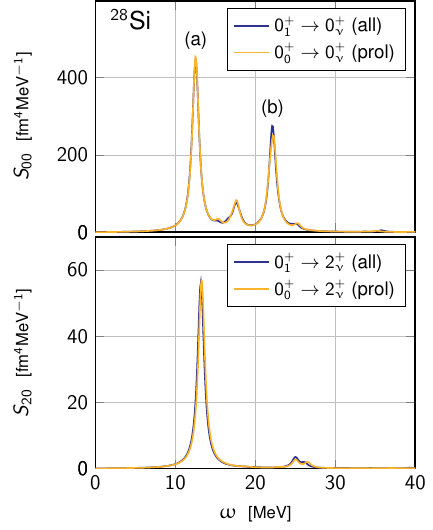}
    \caption{Monopole (top panel) and quadrupole (bottom panel) response of the $0^+$ prolate-shape isomer in $^{28}$Si. The yellow (blue) curve refers to the PGCM calculation limited to prolate (including both oblate and prolate) configurations displayed in Fig.~\ref{fig:PES_Si28}. Excitation energies are measured relatively to the $0^+$ prolate-shape isomer.}
    \label{fig:spectra_Si28_pro}
\end{figure}

The PGCM responses computed with respect to the first excited $0^+$ state corresponding to the prolate-shape isomer are shown in Fig.~\ref{fig:spectra_Si28_pro}. The calculation involving only prolate configurations and the one adding oblate configurations deliver identical results. The decoupling of the two wells is thus even more pronounced than for the ground-state response.  

The monopole response displays a typical structure with two well-separated main peaks, plus a small peak in between. The quadrupole response is essentially concentrated in one peak whose energy is very close to the low-energy component of the GMR. The strength of the GQR is much larger than for the ground-state response.

\begin{figure}
    \centering
    \includegraphics[width=\columnwidth]{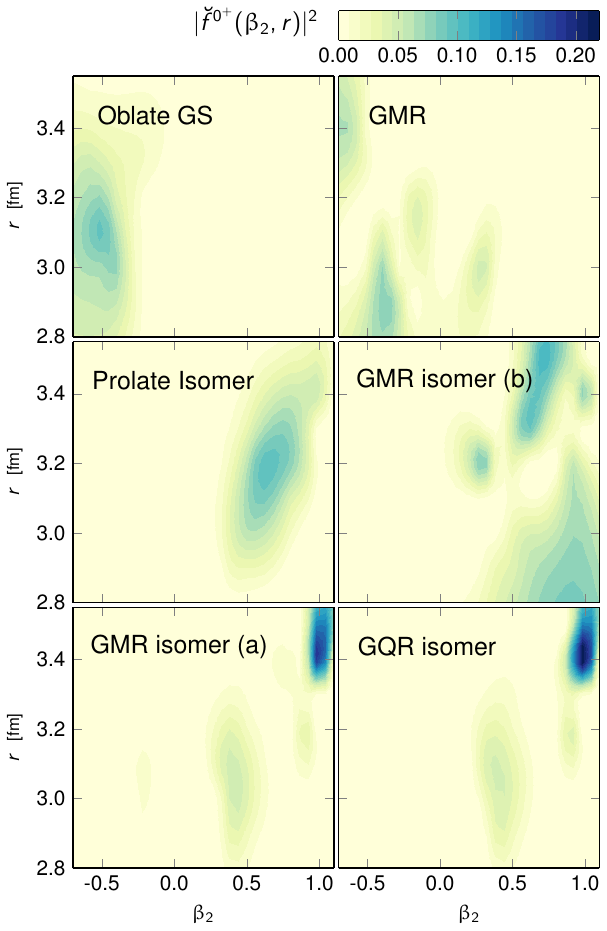}
    \caption{Intrinsic PGCM collective wave-functions in the $(r,\beta_2)$ plane of several states of interest (labels refer to Fig.~\ref{fig:spectra_Si28_pro}) in $^{28}$Si: (top left) $0^+$ ground state $0^+$, (top right) GMR built on the ground state, (middle left) $0^+$ shape isomer, (middle right) $0^+$ state (b) corresponding to the high-energy component of the GMR built on the prolate-shape isomer, (bottom left) $0^+$ state (a)  corresponding to the high-energy component of the GMR built on  the prolate-shape isomer, (bottom right) $2^+$ state corresponding to the GQR built on top of the prolate-shape isomer.}
    \label{fig:Si28_WF}
\end{figure}

\subsection{Nature of the excitations}

The GMR at $18$~MeV built on the oblate ground state is the manifestation, as previously showed in $^{46}$Ti, of a standard breathing mode, i.e. a radial vibration on top of the ground-state configuration. This is seen by comparing the intrinsic PGCM collective wave-function of the GMR in Fig.~\ref{fig:Si28_WF} (top right) to the ground wave-function one (top left). Similarly, the high-energy peak (b) in Fig.~\ref{fig:spectra_Si28_pro} is a standard breathing mode built on the prolate-shape isomer. This is again seen by comparing its intrinsic collective wave-function in the middle panel of Fig.~\ref{fig:Si28_WF} to the one of the prolate-shape isomer (middle left).

\begin{figure}
    \centering
    \includegraphics[scale=0.97,right]{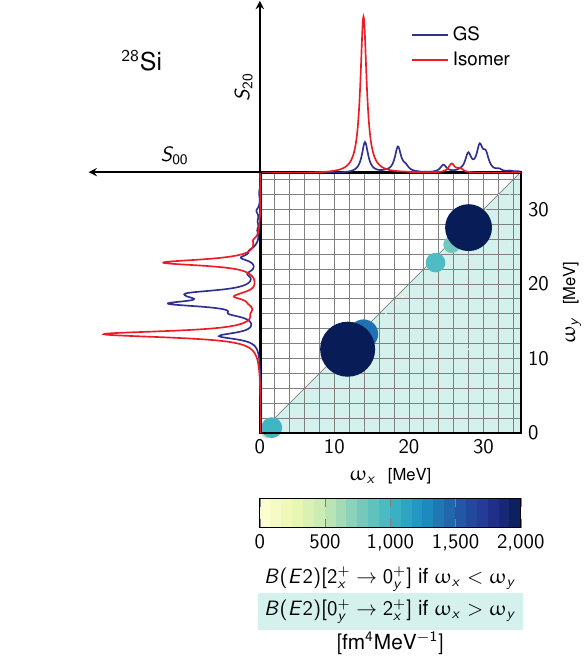}
    \caption{Upward $B(E2)$ transitions between all $0^+$ and $2^+$ states issued from the two-dimensional PGCM calculation of $^{28}$Si. The dots are located at the intersection of the energies between $0^+$ (vertical axis) and $2^+$ (horizontal axis) states. The monopole (quadrupole) response of both the ground state and the shape isomer are shown for comparison on the vertical (horizontal) axis. All excitation energies are measured relative to the ground state. The size of the dots and the colour coding both reflect the magnitude of the transition.}
    \label{fig:Si28_coupling}
\end{figure}

The low-energy  component (a) at $12.2$~MeV (Fig.~\ref{fig:spectra_Si28_pro}) of the GMR built on the prolate-shape isomer is of different nature. As visible from the bottom panel of Fig.~\ref{fig:Si28_WF}, it combines a vibration along $r$ with a vibration along $\beta_2$. Its intrinsic collective wave-function is identical to the one of the corresponding $2^+$ peak, namely the GQR (bottom right). As discussed for $^{46}$Ti, these two states originate from the same intrinsic vibrational state resulting from the GMR-GQR coupling mechanism. Their slightly different excitation energy, i.e. $12.6$~MeV for the $0^+$ state and $13.2$~MeV for the $2^+$, results from the further coupling to the rotational motion captured via the angular momentum projection on $J^\pi=0^+$ and $2^+$, respectively.

\begin{figure*}
    \centering
    \includegraphics[width=\textwidth]{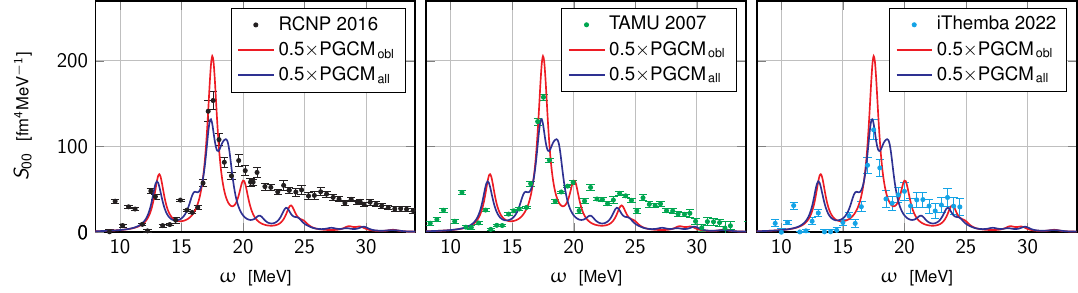}
    \caption{Comparison of PGCM results with experimental monopole responses  in $^{28}$Si from Refs.~\cite{Peach16a,Youngblood07a,Bahini22a} (left, centre and right panel, respectively). Results from the PGCM calculation including only oblate configurations and the PGCM calculation including all configurations are both shown. Theoretical results are multiplied by a 0.5 factor.}
    \label{fig:spectra_Si28_exp}
\end{figure*}

The prolate-shape isomer in $^{28}$Si happens to be an excellent case to corroborate the interpretation of this coupling thanks to the magnitude of the quadrupole strength and to the clean separation of the monopole peaks. This is confirmed via Fig.~\ref{fig:Si28_coupling} displaying upward $B(E2)$ transition probabilities between all $2^+$ and $0^+$ states. A large $B(E2)$ indeed links the GQR and the lower-energy peak of the GMR built on the prolate-shape isomer. The corresponding transition amplitude  $|\braket{0^+_{\!_\text{(a))}}|Q_{20}|2^+_{\!_\text{\;GQR}}}|^2=1408$ fm$^4$MeV$^{-1}$ is of the same order as $|\braket{2^+_{\!_\text{\;GQR}}|Q_{20}|2^+_{\!_\text{\;GQR}}}|^2=2229$ fm$^4$MeV$^{-1}$, the quadrupole moment of the GQR.

\subsection{Comparison to experiment}
\label{sec:Si28_exp}

The theoretical monopole response of the oblate ground state is compared in Fig.~\ref{fig:spectra_Si28_exp} with experimental data from Refs.~\cite{Youngblood07a,Peach16a,Bahini22a}. Results from the PGCM calculation including only oblate HFB states or including both oblate and prolate configurations are reported.  The absolute value of the strength, subject to an empirical smearing associated with a Lorentzian function of width $\Gamma=0.5$~MeV (see Paper I), is further multiplied by a factor $0.5$. Given that the normalisation constant of the experimental data itself carries some uncertainty, a meaningful comparison on an absolute scale is anyway currently not possible.

Overall, present PGCM results are in good agreement with available experimental data. In particular, the positioning of the main GMR peak is accurately reproduced. The structure at $13.0$~MeV associated to the GMR-GQR coupling mechanism is in convincing correspondence with structures in the experimental strength function. 
Nevertheless, the fragmentation testified by the three sets of data, even if not in a univocal way, fails to be described by the PGCM calculation. In Ref.~\cite{Jenkins12a}, such states were suggested to be potential band-heads for super-deformed bands.

Eventually, the high-energy part of the spectrum is also qualitatively reproduced, with the exception of the data from Ref.~\cite{Peach16a} (left plot) that is not consistent with the other two sets in that respect.

\begin{figure}[b]
    \centering
    \includegraphics[width=\columnwidth]{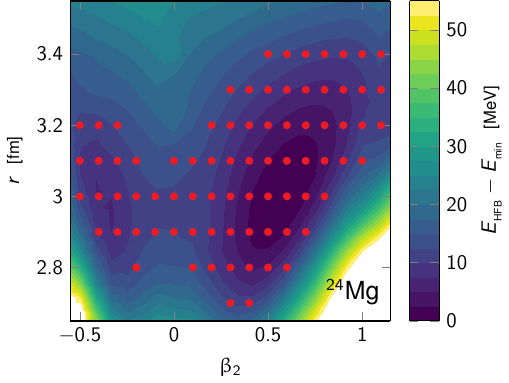}
    \caption{HFB total energy surface for $^{24}$Mg in the $(\beta_2,r)$ plane. The red dots correspond to the configurations included in the subsequent 2D PGCM calculation.}
    \label{fig:PES_Mg24}
\end{figure}

\section{Highly fragmented response of $^{24}$Mg}
\label{sec:Mg24_resphys}

Several physical aspects of the GMR in deformed systems were discussed in previous sections. A particular attention was dedicated to understanding the GMR-GQR coupling mechanism in $^{46}$Ti and $^{28}$Si that is well accounted for within the PGCM. Furthermore, the PGCM was shown to be well suited to disentangle the combined roles (i) of the {\it static} quadrupole deformation and of its {\it fluctuations} along with the roles of (ii)  the coupling of radial and quadrupole fluctuations together with the coupling to rotational motion.  Some nuclei, however, display more complex responses whose structure cannot be easily related to the combined effects of such basic ingredients. This is the case, for instance, of $^{24}$Mg, which is addressed in the present section.

%The HFB energy, deformation and rms radius are displayed in Fig.~\ref{fig:HFB_Mg24} for different model space parameters. The optimal parameter $\hbar\omega$=12~MeV allows a fast convergence with respect to the dimensions of the model space. Calculations are thus performed for $e_{\!_{\!\text{max}}}=$10.

The 2D HFB TES of $^{24}$Mg shown in Fig.~\ref{fig:PES_Mg24} displays a well-defined prolate minimum  at large deformation ($\beta_2=0.56$, see also Table~\ref{tab:Mg24_prop}) accompanied by a secondary oblate minimum about 7~MeV higher in energy. The HFB vacua included in the subsequent PGCM calculation (red dots) cover both prolate and oblate configurations up to $15$~MeV excitation energy above the prolate minimum.

\subsection{Ground-state rotational band}

\begin{figure}
	\centering
	\includegraphics[width=\columnwidth]{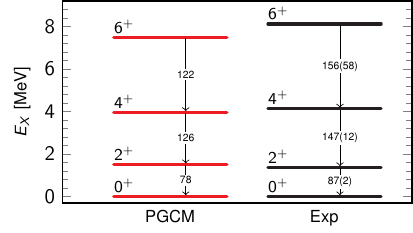}
	\caption{Ground-state rotational band of $^{24}$Mg. Excitation energies and $B(E2)$ values (in  fm$^4$MeV$^{-1}$) from 2D PGCM calculation are shown and compared to experimental data~\cite{Beck09a}.}
	\label{fig:LL_Mg24}
\end{figure}

The ground-state rotational band displayed in Fig.~\ref{fig:LL_Mg24} is shown to be in excellent agreement with experimental data, both regarding excitation energies and (downward) $B(E2)$ values. This gives confidence that the static deformation and the collective character of low-lying states are well captured by the PGCM calculation.
\begin{table}
    \centering
    \begin{tabular}{lcr}
    	\hline
        $E_{\;\text{HFB}}$ [MeV] & $r$ [fm] & $\beta_2$ \\
        \hline
        -110.52 & 2.997 & 0.56 \\ 
        \hline
    \end{tabular}
    \caption{Total energy, rms radius and quadrupole deformation $\beta_2$ of the HFB minimum of the $^{24}$Mg TES from Fig.~\ref{fig:PES_Mg24}.}
    \label{tab:Mg24_prop}
\end{table}

\begin{figure}
    \centering
    \includegraphics[width=\columnwidth]{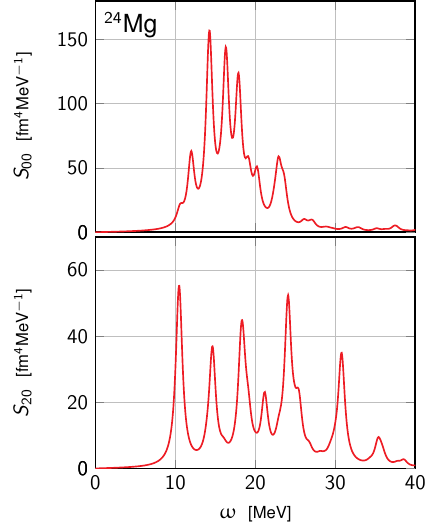}
    \caption{Monopole (top) and quadrupole (bottom) response of $^{24}$Mg from 2D PGCM calculation.}
    \label{fig:spectra_Mg24}
\end{figure}

\begin{figure}
    \centering
    \includegraphics[scale=0.97,right]{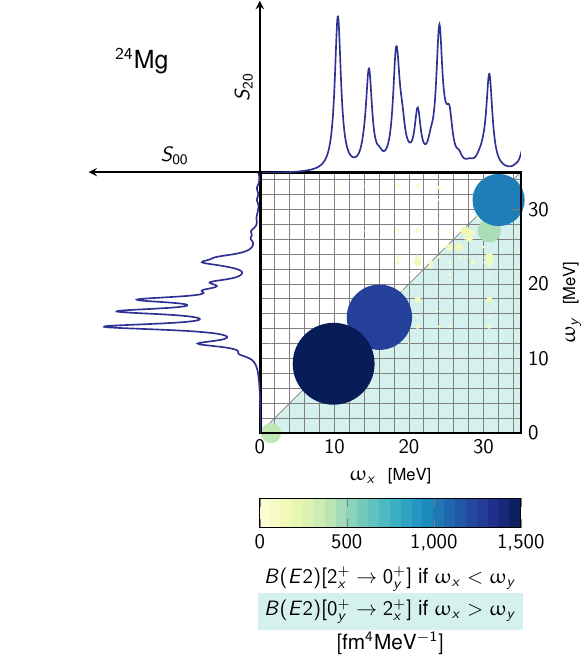}
    \caption{Same as Fig.~\ref{fig:Ti46_PP_coupling} for $^{24}$Mg.}
    \label{fig:coupling_Mg24}
\end{figure}
\begin{figure*}
    \centering
    \includegraphics[width=\textwidth]{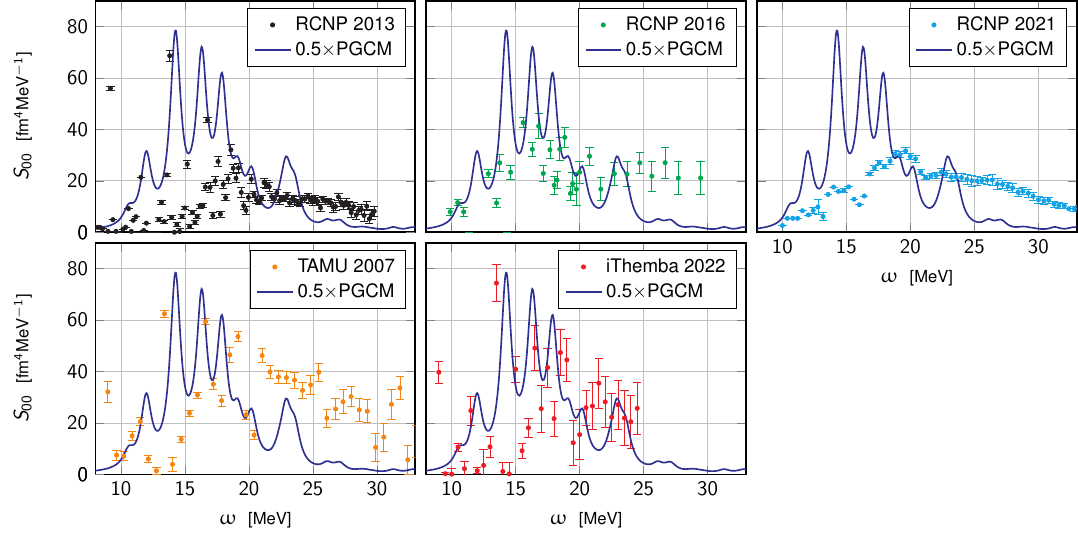}
    \caption{Comparison of the PGCM monopole response with experimental data in $^{24}$Mg from Refs.~\cite{Kawabata13a} (top left), \cite{Gupta15a,Gupta16a} (top centre), \cite{Zamora21a} (top right), \cite{Youngblood09a} (bottom left) and \cite{Bahini22a} (bottom centre). When compared to experimental data, PGCM results are convoluted with a Lorentzian function of width $\Gamma=$0.5 MeV. Theoretical results are multiplied by a 0.5 factor.}
    \label{fig:spectra_Mg24_exp}
\end{figure*}

\subsection{Monopole and quadrupole responses}

The PGCM monopole and quadrupole responses  are displayed in Fig.~\ref{fig:spectra_Mg24}. Differently from previously discussed cases, both responses are extremely fragmented\footnote{The numerical stability of the response has been tested with respect to the linear independence of the norm eigenvalues. The convergence with respect to the set of selected HFB points can affect the fine positioning or the height of the peaks but cannot change the responses significantly; see Paper I for details.}, so that no easy correspondence between the two spectra can be found. In previous examples the GMR could be decomposed into a pure breathing mode and a lower-energy state originating from the coupling with the GQR, accompanied by minor structures. Here, many states carry relatively similar strengths, so that the identification of a dominant mode is more difficult. 

Upward $B(E2)$ transition probabilities between all $2^+$ and $0^+$ states are presented in Fig.~\ref{fig:coupling_Mg24}. They do help identifying the coupling between the $0^+$ and $2^+$ resonant structures near $14.5$~MeV that indeed relate to the same intrinsic collective state. However, no other significant $B(E2)$ transition between resonant structures emerges from the calculation\footnote{The strong $B(E2)$ seen at about $9$~MeV near the diagonal is associated with the $2^+$ rotation on top of the oblate-shape isomer.}.

\subsection{Comparison to experiment}

The PGCM monopole response in $^{24}$Mg is now compared to five sets of experimental data~\cite{Gupta15a,Gupta16a,Kawabata13a,Zamora21a,Youngblood09a} in Fig.~\ref{fig:spectra_Mg24_exp}. As before, a $0.5$ multiplicative factor is used on top of the smearing performed via a Lorentzian function of width $\Gamma=$0.5 MeV. Given the very fragmented nature of the PGCM spectrum, a larger value of $\Gamma$ results into a single very broad response %as also shown in Fig.~\ref{fig:spectra_Mg24_exp} (bottom right) 
and would not lead to any meaningful comparison with experimental data.

The experimental monopole responses are, at best, in weak agreement and, for some of them, essentially incompatible. This is true not only when comparing results from different facilities, but also when successive campaigns from the same infrastructure (RCNP) are considered. Still, except for the results from Ref.~\cite{Zamora21a} (top right), all data sets deliver highly fragmented monopole responses that are not inconsistent with the PGCM prediction. However, no critical comparison between theory and experiment can be presently conducted; a situation that calls for novel experimental investigations. 

\begin{table*}
    \centering
    %\begin{adjustbox}{width=1\textwidth}
    \begin{tabular}{l|rrr|ccc|rr}
    	\hline
    	\hline\vspace{-0.4cm}\\
        ~ & \multicolumn{3}{c|}{$E_{\;\text{GS}}$ [MeV]} & \multicolumn{3}{c|}{$r$ [fm]} & \multicolumn{2}{c}{$\beta_2$} \\
        ~ & HFB & GCM & PGCM & HFB & GCM & PGCM & HFB & GCM \\
        \hline
        $^{16}$O & -79.55 & -79.68 & -79.92 & 2.580 & 2.581 & 2.583 &  0.00 & 0.00 \\ 
        $^{24}$Mg & -110.52 & -111.16 & -116.03 & 2.997 & 3.006 & 3.011 &  0.56 & 0.57 \\ 
        $^{28}$Si & -135.89 & -136.11 & -140.95 & 3.107 & 3.109 & 3.116 &  -0.45 & -0.45 \\ 
        $^{28}$Si$_{\;\text{iso}}$ & -134.41 & -135.05 & -140.23 & 3.168 & 3.179 & 3.199 & 0.60 & 0.61 \\ 
        $^{46}$Ti & -237.30 & -237.64 & -240.69 & 3.474 & 3.474 & 3.483 &  0.27 & 0.26 \\ 
        \hline
        \hline
    \end{tabular}
    %\end{adjustbox}
    \caption{Total energy, rms radius and quadrupole deformation $\beta_2$ of the HFB minima employed in QFAM calculations. Results from (P)GCM calculations are also reported for comparison.} %, as well as experimental values for charge radii~\cite{Fricke95a}
    \label{tab:GS_prop}
\end{table*}

\begin{figure*}
    \centering
    \includegraphics[width=\textwidth]{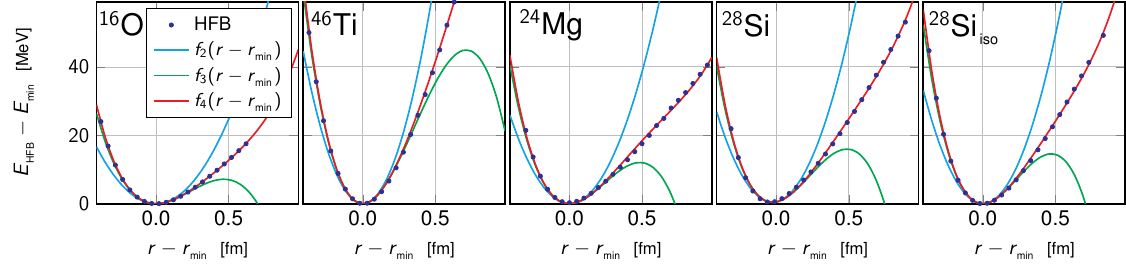}
    \caption{One-dimensional HFB total energy surface  for fixed $\beta_2=\beta_{2_{\;\text{min}}}$. Polynomial fits (see Eq.~\eqref{eq:fit_funct}) are also reported for different values of $k$. The fitting parameters are reported in Tab.~\ref{tab:fit_param}.}
    \label{fig:pes_1d_fit}
\end{figure*}

\section{Comparison to QRPA}
\label{sec:QFAM_res}

In this section, PGCM results discussed above are compared to those obtained from QRPA calculations. \textit{Ab initio} QRPA calculations are performed via the finite amplitude method (FAM)~\cite{Beaujeault23a}. The FAM was originally developed within an EDF setting, first addressing closed-shell systems~\cite{Nakatsukasa07a}. It was later extended to open-shell systems via the generalisation to the quasiparticles realm~\cite{Avogadro11a}, producing the so-called quasiparticle finite amplitude method (QFAM). The (Q)FAM approach is strictly equivalent to the traditional (Q)RPA but presents some numerical advantages. An \textit{ab initio} implementation of QFAM applicable to axially- and triaxially-deformed systems was recently achieved\footnote{While the \textit{ab initio} QFAM code can access any multipolarity, its use is presently limited to computing monopole responses.}~\cite{Beaujeault23a}. 

The comparison between QRPA and PGCM calculations provides useful benchmarks, keeping in mind that they are not expected to provide the same results. First, QRPA can be proven to be the harmonic approximation to the {\it most general} GCM~\cite{Porro:2023bef} such that it may miss anharmonic effects. On the other hand, practical GCM calculations cannot handle more than a few collective coordinates, e.g. two in the present case, such that the complete manifold of Bogoliubov states is not visited by the PGCM ansatz, even within the harmonic limits. While the QRPA explores that manifold in all directions allowed by the symmetry restrictions of the specific implementation, it does so in a way that is limited to quadratic fluctuations around the HFB minimum. The compromise between the harmonic approximation and the systematic exploration of the generator coordinates space is here investigated via the explicit comparison between QRPA and PGCM. The second major difference between QRPA and PGCM is the exact restoration of good symmetries achieved by the latter\footnote{While steps towards the explicit inclusion of symmetry restoration within QRPA have been taken recently~\cite{Porro:2023yto}, an realistic implementation of the full-fledged projected (Q)RPA formalism~\cite{FedRi85} is still lacking. }. The impact of restoring good angular-momentum is presently disentangled by displaying results from both PGCM and GCM calculations, the latter being performed using the same numerical setting as the one presently employed in PGCM calculations\footnote{Particle-number projection is included by default, such that only the impact of angular-momentum projection is presently addressed.}. 

\subsection{Ground-state properties}

Results presented in Secs.~\ref{sec:O16_resphys}, \ref{sec:Ti46_resphys}, \ref{sec:Si28_resphys} and \ref{sec:Mg24_resphys} are compared to results from GCM and QFAM calculations performed in a consistent setting. Ground-state properties at HFB, GCM and PGCM levels are listed in Tab.~\ref{tab:GS_prop}.

Ground-state energies are not significantly lowered by the configuration mixing included in the GCM, i.e. by typically less than 1~MeV. Angular momentum projection allows instead to gain up to an additional $6$~MeV in doubly open-shell nuclei. Radii are essentially not affected by static correlations, which makes an accurate reproduction at the HFB level a crucial step for the subsequent (P)GCM calculation. The intrinsic quadrupole deformation $\beta_2$ is also stable going from HFB to GCM\footnote{The PGCM ground states carrying angular momentum $J~=~0$, their static quadrupole moment is null by definition. It is however possible to compute an average intrinsic deformation from the associated intrinsic collective wave-function. Such quantity is not displayed here.}.

\subsection{Anharmonic effects}
\label{sec:anharm_corr}

In order to quantify anharmonic effects on the GMR, an analysis of beyond-quadratic corrections based on a simple model is now proposed. Considering the 1D cut of the PES at fixed $\beta_2=\beta_{2_{\;\text{min}}}$, the class of polynomials
\begin{equation}
    f_k(x)\equiv\sum_{l=2}^ka_lx^l \, ,
    \label{eq:fit_funct}
\end{equation}
is employed to fit the energy profile as a function of $r-r_{\text{min}}$. In all nuclei, fourth-order polynomials are found to accurately describe the observed radial behaviour, as shown in Fig.~\ref{fig:pes_1d_fit} where the quadratic, cubic and quartic fitting functions are also displayed for comparison. The corresponding fitting parameters are reported in Tab.~\ref{tab:fit_param}.

\begin{table}[]
    \centering
    \begin{tabular}{lrrr}
    	\hline
    	\hline\vspace{-0.4cm}\\
        ~ & $a_2$ & $a_3$ & $a_4$ \\ 
        ~ & [MeV fm$^{-2}$] & [MeV fm$^{-3}$]& [MeV fm$^{-4}$]\\ 
        \hline
        $^{16}$O & 98.91 & -141.37 & 90.10  \\ 
        $^{24}$Mg & 154.18 & -212.12 & 102.23  \\ 
        $^{28}$Si  & 178.09 & -231.77 & 116.75 \\ 
        $^{28}$Si$_{\;\text{iso}}$ & 197.10 & -278.76 & 155.20 \\ 
        $^{46}$Ti & 265.44 & -248.26 & 100.20 \\ 
        \hline
        \hline
    \end{tabular}
    \caption{Fitting parameters relative to the functions plotted in Fig.~\ref{fig:pes_1d_fit}. The function from Eq.~\eqref{eq:fit_funct} with $k=4$ was used.}
    \label{tab:fit_param}
\end{table}

\begin{table*}[]
    \centering
    \begin{tabular}{lrrrrc}
    	\hline
    	\hline\vspace{-0.4cm}\\
        ~ & $a_2$ [MeV fm$^{-2}$] & $\hbar\omega$ [MeV] & $E^{\;\!^\text{GMR}}_{\;\!_\text{QFAM}}$ [MeV] & $\Delta E_{\;\text{sph}}$ [MeV]& $|\beta_2|$ \\ 
        \hline
        $^{16}$O & 98.91 & 22.62 & 22.70 & -0.08 & 0.00 \\ 
        $^{24}$Mg & 154.18 & 23.06 & 17.10 & 5.96 & 0.56 \\ 
        $^{28}$Si  & 178.09 & 22.94 & 17.90 & 5.04 & 0.45 \\ 
        $^{28}$Si$_{\;\text{iso}}$ & 197.10 & 24.14 & 17.30 & 6.84 & 0.60 \\ 
        $^{46}$Ti & 265.44 & 21.85 & 19.80 & 2.05 & 0.27 \\ 
        \hline
        \hline
    \end{tabular}
    \caption{Eigen-frequencies extrapolated from the harmonic approximation to one-dimensional energy surfaces at fixed $\beta_2=\beta_{2_{\;\text{min}}}$ from Fig.~\ref{fig:pes_1d_fit}. The GMR energy from QFAM calculations are also reported for comparison.}
    \label{tab:def_harm}
\end{table*}

Based on the quadratic fit, the eigen-frequency $\hbar\omega$ of the associated quantum harmonic oscillator model is evaluated. The extracted values are tabulated in Tab.~\ref{tab:def_harm} and compared to the GMR energy $E^{\;\!^\text{GMR}}_{\;\!_\text{QFAM}}$ from QFAM calculations discussed later on. The energy difference 
\begin{equation}
    \Delta E_{\;\text{sph}}\equiv \hbar\omega-E^{\;\!^\text{GMR}}_{\;\!_\text{QFAM}}
\end{equation}
quantifies the departure from the harmonic limit.

\begin{figure}
    \centering
    \includegraphics[width=0.47\columnwidth,valign=t]{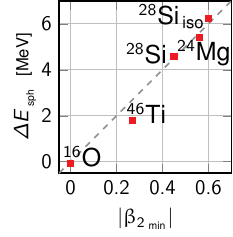}
    \hfill
    \includegraphics[width=0.51\columnwidth,valign=t]{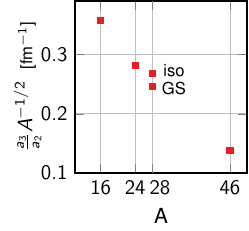}
    \caption{\textbf{Left:} Effects of deformation on the GMR positioning in QFAM. \textbf{Right:} beyond-quadratic fitting parameter as a function of the mass number.}
    \label{fig:def_harm}
\end{figure}

A quantitative agreement between this simple model and QFAM results is only observed for $^{16}$O. This can be expected given that extracting the eigen-frequency from the 1D radial energy profile implicitly assumes that the monopole resonance is decoupled from other oscillating modes. This hypothesis is well verified for intrinsically-spherical systems such as $^{16}$O. However, as discussed at length above, the GMR couples to the quadrupole mode in intrinsically-deformed systems. As visible from the left panel of Fig.~\ref{fig:def_harm}, $\Delta E_{\;\text{sph}}$ is indeed proportional to the intrinsic deformation of the system \cite{Nishizaki85a}, which tends to support the assumption that  a coupled anisotropic harmonic oscillator model should be considered instead for such nuclei.

\begin{table*}[t]
    \centering
    \begin{tabular}{lrrrrrrrr}
    	\hline
    	\hline\vspace{-0.4cm}\\
        ~ & $\hbar\omega$ & $E_0^{(2,3)}$ & $E_0^{(1,4)}$ & $E_0^{(2,4)}$ & $E_1^{(2,3)}$  & $E_1^{(1,4)}$ & $E_1^{(2,4)}$ & $\hbar\omega^{(1+2)}$ \\
        \hline
        $^{16}$O & 22.618 & -1.816 & 0.883 & -0.161 & -11.720 & 4.416 & -1.265 & 15.143 \\ 
        $^{24}$Mg & 23.056 & -1.122 & 0.429 & -0.037 & -7.240 & 2.143 & -0.292 & 18.397 \\ 
        $^{28}$Si  & 22.941 & -0.860 & 0.363 & -0.027 & -5.553 & 1.816 & -0.211 & 19.518 \\ 
		$^{28}$Si$_{\;\text{iso}}$ & 24.135 & -1.016 & 0.436 & -0.037 & -6.558 & 2.182 & -0.289 & 20.086 \\ 
        $^{46}$Ti & 21.851 & -0.270 & 0.127 & -0.003 & -1.746 & 0.637 & -0.027 & 20.862 \\ 
		\hline
		\hline
    \end{tabular}
    \caption{First- and second-order perturbative corrections to the harmonic-oscillator ($r$) eigen-energies for the ground state and the first excited state (resonance) associated to the fitting parameters from Tab.~\ref{tab:fit_param}. All values are expressed in MeV units. See \ref{app:pert_corr} for details on labels.}
    \label{tab:corr_mono}
\end{table*}

While the simple 1D model presently employed does not allow a proper quantification of anharmonic effects in deformed systems, the magnitude of the fitting coefficients beyond the quadratic term is an indicator of their importance. The ratios of the cubic coefficients with respect to the harmonic coefficient are thus plotted in the right panel of Fig.~\ref{fig:def_harm} as a function of the nuclear mass, rescaled by a factor $A^{-1/2}$ in order to remove trivial $A$ dependencies\footnote{The rms radius, which is used as the fitting variable in Eq.~\eqref{eq:fit_funct}, is defined over a generic state $\ket{\Phi}$ as
\begin{equation}
    r_{rms}\equiv\sqrt{\frac{\braket{\Phi|r^2|\Phi}}{A}}\,.
\end{equation}}. The ratio decreases with the nuclear mass. In this respect, $^{16}$O is the least harmonic nucleus, whereas $^{46}$Ti is the most harmonic one. 
%This is coherent with the Pauli principle violation (which is implicit in the Quasi Boson Approximation determining the harmonic approximation, see Ref.~\cite{} for details) being less severe for more collective states and, thus, in heavier systems~\cite{Ring80a}. 
Globally, deviations between QFAM and (P)GCM are thus expected to be more prominent in light nuclei.

Given the magnitude of the coefficients ratios, the validity of a perturbative analysis of beyond-quadratic corrections is questionable. Still, first- and second-order perturbative corrections to the ground- and  first-excited-state (resonance) energies are provided in Tab.~\ref{tab:corr_mono} based on the analytical evaluation laid down in \ref{app:pert_corr}. As proven in \ref{app:pert_corr}, first-order cubic corrections $E_n^{(1,3)}$ vanish identically. Second-order cubic corrections $E_n^{(2,3)}$ are larger in light systems, which is symptomatic of significant anharmonic effects. Possibly, third-order corrections are necessary in order to converge the perturbative series. Quartic corrections are less important for all observed cases, and a clear convergent trend is visible going from first ($E_n^{(1,4)}$) to second order ($E_n^{(2,4)}$).

While a quantitative  understanding of the energy correction from anharmonic effects requires an analysis including the coupling between $r$ and $\beta_2$ generator coordinates, the uncoupled evaluation of perturbative corrections to the harmonic picture proposed above already provides a qualitative tool to explain possible differences between QFAM and (P)GCM calculations.

\begin{figure}
    \centering
    \includegraphics[width=\columnwidth]{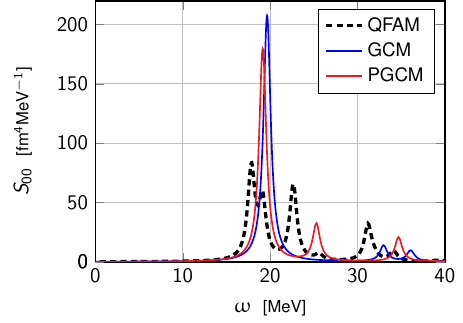}
    \caption{Monopole response in $^{16}$O from QFAM and (P)GCM calculations.} 
    \label{fig:O16_QFAM}
\end{figure}

\begin{figure}
    \centering
    \includegraphics[width=\columnwidth]{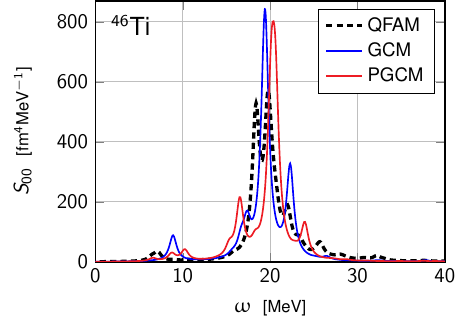}
    \caption{Monopole response in $^{46}$Ti from QFAM and (P)GCM calculations.}
    \label{fig:spectra_Ti46_QFAM}
\end{figure}

\subsection{Comparison in $^{16}$O}

The monopole responses of $^{16}$O from QFAM and (P)GCM calculations are shown in Fig.~\ref{fig:O16_QFAM}. QFAM displays a more fragmented strength function than  (P)GCM. The peak most naturally associated with the GMR is the one at $22.7$~MeV, since it is very close to the first excitation in the pure 1D harmonic oscillator model discussed above (see Tab.~\ref{tab:def_harm}). The GMR predicted by GCM\footnote{The GMR predicted by PGCM is only $500$~keV below the GCM one. Furthermore, the associated transition amplitudes (before their convolution with a Lorentzian function) differ only by 40 fm$^4$MeV$^{-1}$. Overall, the two monopole responses are almost identical. This demonstrates the small effect of the angular momentum projection, which is itself due to the negligible weight carried by intrinsically-deformed HFB configurations in the $0^+$ PGCM states.} appears $\sim$3.5 MeV below. The perturbative estimate of anharmonic corrections predicts a shift down by $7.5$~MeV (see Tab.~\ref{tab:corr_mono}). While this number is definitely overestimated due to the lack of validity of the perturbative approach, especially in light nuclei, it qualitatively explains the lower location of the GMR in (P)GCM calculations. 

The QFAM response displays two lower-lying peaks that are absent from (P)GCM results. These states cannot result from anharmonic effects in the $(r, \beta_2)$ plane, not captured by QRPA calculations, and must thus originate from the coupling of the radial motion with other generator coordinates than $\beta_2$  not included in the presently employed (P)GCM ansatz.

%Eventually, different smearing parameters are explored. Increasing $\Gamma$ to $1.5$~MeV or $2.0$~MeV, the difference between PGCM and QFAM results becomes less and less pronounced, as visible in Fig.~\ref{fig:O16_QFAM}. In such cases, a global agreement between QFAM and PGCM spectra is observed.

\subsection{Comparison in $^{46}$Ti}

The monopole responses in $^{46}$Ti are compared in Fig.~\ref{fig:spectra_Ti46_QFAM}. The three methods agree on the position of the high-energy component of the GMR, i.e. at 19.8~MeV for QFAM, 20.4~MeV for PGCM and 19.4~MeV for GCM calculations. The consistent description of the GMR testifies the validity of the harmonic approximation for such a  breathing-mode-like peak in $^{46}$Ti.

The low-energy QFAM component at 18.4~MeV is associated to the quadrupole peak at the same energy (not shown) and results from the GMR-GQR coupling mechanism. Its energy is significantly higher than the corresponding $0^+$ state at $16.5$~MeV in PGCM whereas the GCM locates this peak in between. Somewhat naively, this picture leads to attributing half of the lowering of that GMR-GQR coupling peak in PGCM compared to QFAM to anharmonic effects and another half to angular momentum projection effects. Eventually, the PGCM value must be compared to the experimental one of $16.8$~MeV obtained from the centroid of the fitted Gaussian in Ref.~\cite{Tokimoto06a}.

\begin{figure}
    \centering
    \includegraphics[width=\columnwidth]{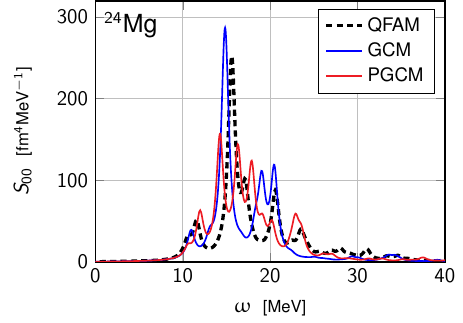}
    \caption{Monopole response in $^{24}$Mg from QFAM and (P)GCM calculations.}
    \label{fig:spectra_Mg24_QFAM}
\end{figure}

%The low-energy quadrupole response is different for the three methods. The low-energy quadrupole strength arising from PGCM calculations in $^{46}$Ti was already addressed in Sec.~\ref{sec:Ti46_res_quad}. The GCM quadrupole response in the 5-15~MeV region is different from its projected counterpart. One strong transition is observed at $\sim$9~MeV instead of many fragmented peaks. However, the associated physical content is similar. A similar response is also observed, even if at lower energies, for the QFAM quadrupole response. The collective GCM wave-function of the 9~MeV quadrupole peak is compared to one of the $2^+$ PGCM states at $\sim10$~MeV in Fig.~\ref{fig:WF_Ti46_UNP}. The strong quadrupole vibration is visible in both cases. %Due to its energy position, this transition is postulated to be associated to excitations of non-collective type.

%Overall, a better agreement between PGCM and QFAM calculations is observed in $^{46}$Ti for the monopole rather than for the quadrupole response. The GQR amplitude and the corresponding manifestation in the GMR is larger in QFAM than in (P)GCM, and the associated GMR splitting is underestimated with respect to (P)GCM calculations.

\subsection{Comparison in $^{24}$Mg}

The monopole responses of $^{24}$Mg are compared in Fig.~\ref{fig:spectra_Mg24_QFAM}. A qualitative agreement is observed between GCM and QFAM. The  GMR peak in GCM  (14.8 MeV) is slightly lower than the QFAM value (15.6 MeV). The small energy difference may be attributed to anharmonic effects, even though the corresponding shift was estimated to be much larger ($\sim$5~MeV) based on (a questionable) perturbative evaluation (see Tab.~\ref{tab:corr_mono}). Overall a similar two-peak structure is predicted in both calculations. Including angular momentum projection modifies the response significantly, the PGCM monopole response being more fragmented. It will be interesting to verify if a similar effect is observed in future projected (Q)RPA calculations~\cite{Porro:2023yto}.

\begin{figure}
    \centering
    \includegraphics[width=\columnwidth]{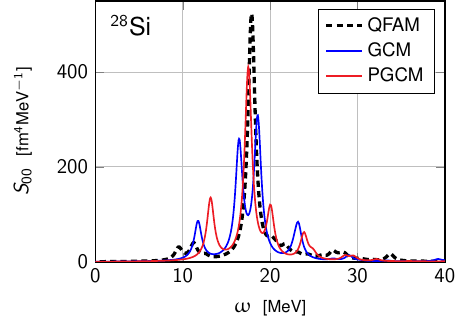}
    \caption{Monopole response of the ground state in $^{28}$Si from QFAM and (P)GCM calculations.}
    \label{fig:spectra_Si28_QFAM}
\end{figure}

\begin{figure}
    \centering
    \includegraphics[width=\columnwidth]{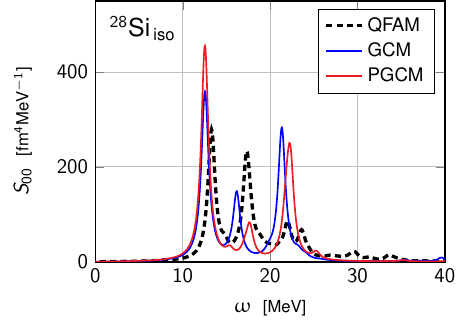}
    \caption{Monopole response of the prolate-shape isomer in $^{28}$Si from QFAM and PGCM calculations.}
    \label{fig:spectra_Si28_iso_QFAM}
\end{figure}

\subsection{Comparison in $^{28}$Si}

The ground-state and the prolate-shape isomer monopole responses in $^{28}$Si are compared  in Figs.~\ref{fig:spectra_Si28_QFAM} and~\ref{fig:spectra_Si28_iso_QFAM}, respectively. For the ground-state response, the breathing mode at $17.5$~MeV  is coherently predicted by QFAM and (P)GCM calculations, knowing that it is split into two peaks in the GCM. Contrarily, the low-energy component at $\sim$13~MeV in (P)GCM calculations is not observed in QFAM. 

%As for the ground-state quadrupole channel, the three methods agree in determining a faint response. This is in line with the very partial exhaustion of the associated EWSR in QFAM and (P)GCM calculations, which is reported in Sec.~\ref{sec:mom_ewsr}. This effect is associated to the exploration of the sole $K=0$ component of the quadrupole resonance: the ground state of $^{28}$Si being strongly oblate, quadrupolar vibrations are expected to be more prominent perpendicularly to the symmetry axis. Such kind of quadrupolar vibrations would demand the inclusion of $K\neq0$ components, requiring triaxiality to be considered. 

The response of the prolate-shape isomer is similarly split into three peaks in QFAM and (P)GCM calculations. The lowest peak at about $13$~MeV unambiguously originates from the GMR-GQR coupling mechanism in all methods. The lowest strength of that state in QFAM is reshuffled into the second peak near $17$~MeV that is more pronounced than in (P)GCM, knowing that the two peaks are closer in QFAM than in PGCM by about $1.12$~MeV. The third peak at about $22$ MeV is associated to the breathing mode  whose collective PGCM wave-function was shown in Fig.~\ref{sec:Si28_resphys}. Surprisingly, the strength of that state carries little strength in QFAM. 

%GCM and QFAM quadrupole spectra display a very similar shape, with GCM being shifted down by roughly 1-2~MeV. This effect is assumed to be produced by anharmonicities based on data from Tab.~\ref{tab:corr_quad}. However, projection largely reduces the strength associated to the GQR in GCM calculations, such that a similar effect shall be expected in projected-QRPA calculations.

\subsection{Discussion}

In most studied cases, QFAM and PGCM monopole strengths are at least in qualitative agreement. The largest differences were observed in the lightest $^{16}$O nucleus, where anharmonic effects captured by the (P)GCM shift the GMR down significantly. In addition, the present 2D (P)GCM calculations miss lower-lying structures accessed via QFAM.

Overall, the angular momentum restoration does not impact the monopole responses too significantly, except in specific systems such as $^{24}$Mg. Furthermore, the impact is typically more important on the quadrupole response. This question is not further elaborated on here given that it constitutes the focus of Paper IV.

\section{Conclusions}
\label{sec:concl}

The present article was dedicated to the application of the PGCM to the \textit{ab initio} determination of the isoscalar GMR in $^{16}$O, $^{46}$Ti, $^{28}$Si and $^{24}$Mg, the last three being doubly open-shell nuclei.

The study demonstrated that PGCM results for sd-shell nuclei are in excellent agreement with available experimental data. In fact, the comparison in $^{28}$Si and $^{24}$Mg illustrated the need for new unambiguous experimental data in this mass region. 

The fact that two-dimensional PGCM calculations account well for the fragmented monopole response of (rather) light doubly open-shell nuclei was shown to be due to their capacity (i) to capture the impact on the position of the breathing mode at play in spherical nuclei of the intrinsic static quadrupole deformation and of its fluctuation, (ii) to describe in a refined way the GMR-GQR coupling mechanism responsible for the appearance of an additional component in the GMR of intrinsically-deformed nuclei and (iii) to seize anharmonic effects that were shown to be significant in light systems.

Along the way, 2D PGCM calculations were employed to extract the monopole response on top of the prolate shape isomer in $^{28}$Si. The signal was shown to be limpid with the emergence of a particularly strong peak associated with the GMR-GQR coupling mechanism. The (challenging) possibility to investigate experimentally the monopole response of shape isomers should be envisioned in the future.

\section*{Acknowledgements}
Calculations were performed by using HPC resources from GENCI-TGCC (Contract No. A0130513012). A.P. was supported by the CEA NUMERICS program, which has received funding from the European Union's Horizon 2020 research and innovation program under the Marie Sk{\l}odowska-Curie grant agreement No 800945. A.P. and R.R. are supported by the Deutsche Forschungsgemeinschaft (DFG, German Research Foundation) – Projektnummer 279384907 – SFB 1245. R.R. acknowledges support though the BMBF Verbundprojekt 05P2021 (ErUM-FSP T07, Contract No. 05P21RDFNB).

\section*{Data Availability Statement}

This manuscript has no associated data or the data will not be deposited.

\appendix

\section{Quantum harmonic oscillator and perturbative corrections}
\label{app:pert_corr}

Perturbative corrections up to second order to the eigen-energies of the one-dimensional quantum harmonic oscillator (QHO) are evaluated for a perturbation containing both a cubic and a quartic term\footnote{The perturbation theory for the $q^4$ correction to the QHO is known to diverge (zero radius of convergence)~\cite{Bender69a,Banks73a,Bender73a}.}. The Hamiltonian is thus given by
\begin{equation}
    H=H_0+ \lambda H_1
\end{equation}
with
\begin{subequations}
    \begin{align}
        H_0&\equiv\frac{p^2}{2m}+\frac{1}{2}m\omega^2q^2\,,\\
        H_1&\equiv-\mu q^3+\xi q^4\,,
    \end{align}
\end{subequations}
representing the unperturbed QHO and the perturbation respectively. The canonical position and momentum coordinates can be expressed in terms of ladder operators as
\begin{subequations}
    \begin{align}
        q&= \sqrt{\frac{\hbar}{2m\omega}}(a^\dagger+a)\,,\label{eq:qop_def}\\
        p&= i\sqrt{\frac{m\omega}{2\hbar}}(a^\dagger-a)\,,
    \end{align}
\end{subequations}
$a^\dagger$ and $a$ being defined through their action on QHO eigenstates as
\begin{subequations}
    \begin{align}
        a^\dagger\ket{n}&\equiv\sqrt{n+1}\ket{n+1}\,,\\
        a\ket{n}&\equiv\sqrt{n}\ket{n-1}\, .
    \end{align}
\end{subequations}
The QHO eigenstates $\ket{n}$ themselves fulfill
\begin{equation}
    H_0\ket{n}=E_n^{(0)}\ket{n}\,.
\end{equation}
with 
\begin{equation}
    E_n^{(0)}\equiv \hbar\omega\Big(n+\frac{1}{2}\Big)\,.
\end{equation}
The matrix elements of the operator $q$ are determined using the representation of Eq.~\eqref{eq:qop_def} and read as
\begin{equation}
    \braket{m|q|n}=\sqrt{\frac{\hbar}{2m\omega}}[\sqrt{n}\delta_{m,n-1}+\sqrt{n+1}\delta_{m,n+1}]\,,
\end{equation}
allowing the computation of the transition probability between neighbouring states
\begin{equation}
	\label{eq:linear_trans_trend}
    |\braket{n-1|q|n}|^2=\frac{\hbar}{2m\omega}n\,.
\end{equation}
Eigen-energy corrections at first and second order in perturbation theory are provided, respectively, by
\begin{subequations}
    \begin{align}
        E^{(1)}_n&=\lambda\braket{n|H_1|n}\,,\\
        E^{(2)}_n&=\lambda^2\sum_{m\neq n}\frac{|\braket{n|H_1|m}|^2}{E^{(0)}_n-E^{(0)}_m}\,.
    \end{align}
\end{subequations}

The cubic and the quartic contributions to the first-order term read as
\begin{subequations}
    \begin{align}
        E_n^{(1,3)}&\equiv-\braket{n|\mu q^3|n}\nonumber\\
        &=-\mu\Lambda^{\frac{3}{2}}\braket{n|(a^\dagger+a)^3|n}\nonumber\\
        &=0\,,\\
        E_n^{(1,4)}&\equiv\braket{n|\xi q^4|n}\nonumber\\
        &=\xi\Lambda^2\braket{n|(a^\dagger+a)^4|n}\nonumber\\
        &=\xi\Lambda^2\,3(2n^2+2n+1)\,,
    \end{align}
\end{subequations}
where
\begin{equation}
    \Lambda\equiv\frac{\hbar}{2m\omega}\,.
\end{equation}
The cubic term does not contribute at first order, since its expansion only provides unequal powers of $a^\dagger$ and $a$. The second-order contributions are computed via tedious but straightforward algebraic derivations delivering
\begin{subequations}
    \begin{align}
        E_n^{(2,3)}&\equiv\sum_{m\neq n}\frac{|\braket{n|\mu q^3|m}|^2}{E^{(0)}_n-E^{(0)}_m}\nonumber\\
        &=\frac{\mu^2}{\hbar\omega}\Lambda^3\sum_{m\neq n}\frac{|\braket{n|(a^\dagger+a)^3|m}|^2}{n-m}\nonumber\\
        &=-\frac{\mu^2}{\hbar\omega}\Lambda^3(30n^2+30n+11)\,,\\
        E_n^{(2,4)}&\equiv\sum_{m\neq n}\frac{|\braket{n|\xi q^4|m}|^2}{E^{(0)}_n-E^{(0)}_m}\nonumber\\
        &=\frac{\xi^2}{\hbar\omega}\Lambda^4\sum_{m\neq n}\frac{|\braket{n|(a^\dagger+a)^4|m}|^2}{n-m}\nonumber\\
        &=-\frac{\xi^2}{\hbar\omega}\Lambda^4(68n^3+102n^2+118n+42)\,.
    \end{align}
\end{subequations}
Eventually, the perturbatively-corrected eigen-energies are defined as
\begin{equation}
	E_n^{(1+2)}\equiv E^{(0)}_n+E_n^{(2,3)}+E_n^{(1,4)}+E_n^{(2,4)}\,,
\end{equation}
and, consequently, the perturbed eigen-frequencies as
\begin{equation}
	\hbar\omega^{(1+2)}\equiv E_n^{(1+2)} - E_0^{(1+2)}\,.
\end{equation}

%\section{Transition amplitudes correction}

%\begin{equation}
%    \ket{\Tilde{n}}=\ket{n}+\lambda\sum_{m\neq n}\ket{m}\frac{\braket{m|H_1|n}}{E_n^{(0)}-E_m^{(0)}}+\mathcal{O}(\lambda^2)
%\end{equation}

%\begin{align}
%    \braket{\Tilde{k}|q|\Tilde{n}}&=\braket{k|q|n}+\lambda\sum_{m\neq n}\frac{\braket{k|q|m}\braket{m|H_1|n}}{E_n^{(0)}-E_m^{(0)}}+\lambda\sum_{m\neq k}\frac{\braket{k|H_1|m}\braket{m|q|n}}{E_k^{(0)}-E_m^{(0)}} +\mathcal{O}(\lambda^2)\nonumber\\
%    &=\braket{k|q|n}+\lambda\sum_{m\neq n}\frac{\braket{k|q|m}\braket{m|H_1|n}}{E_n^{(0)}-E_m^{(0)}}+\lambda\sum_{m\neq k}\frac{\braket{n|q|m}\braket{m|H_1|k}}{E_k^{(0)}-E_m^{(0)}} +\mathcal{O}(\lambda^2)\nonumber\\
%    &=\braket{k|q|n}+\lambda\sum_{m\neq n}\frac{\braket{k|q|m}\braket{m|H_1|n}}{E_n^{(0)}-E_m^{(0)}}+(n\xleftrightarrow[]{} k) +\mathcal{O}(\lambda^2)
%\end{align}

\bibliography{biblio.bib}

\end{document}